\documentclass[prd,twocolumn,notitlepage,nofootinbib,longbibliography,superscriptaddress]{revtex4-1}

\usepackage[english]{babel}
\usepackage{amsmath,amssymb}
\usepackage{graphicx,float,placeins}
\usepackage{comment}
\usepackage{diagbox}
\usepackage{booktabs}
\usepackage[colorlinks=true, allcolors=blue]{hyperref}
\usepackage{textgreek}

\newcommand{\E}[1]{%
    \times 10^{#1}%
}

\begin{document}

\title{Approaches to photon absorption in a Lorentz invariance violation scenario}

\author{J. M. Carmona}
\email{jcarmona@unizar.es}
\affiliation{Departamento de Física Teórica and Centro de Astropartículas y Física de Altas Energías (CAPA), Universidad de Zaragoza, Zaragoza 50009, Spain}

\author{J. L. Cortés}
\affiliation{Departamento de Física Teórica and Centro de Astropartículas y Física de Altas Energías (CAPA), Universidad de Zaragoza, Zaragoza 50009, Spain}

\author{F. Rescic}
\affiliation{Departamento de Física Teórica and Centro de Astropartículas y Física de Altas Energías (CAPA), Universidad de Zaragoza, Zaragoza 50009, Spain}
\affiliation{University of Rijeka, Faculty of Physics, Rijeka 51000, Croatia}

\author{M. A. Reyes}
\affiliation{Departamento de Física Teórica and Centro de Astropartículas y Física de Altas Energías (CAPA), Universidad de Zaragoza, Zaragoza 50009, Spain}

\author{T. Terzi\'c}
\email{tterzic@phy.uniri.hr}
\affiliation{University of Rijeka, Faculty of Physics, Rijeka 51000, Croatia}

\author{F. I. Vrban}
\affiliation{Faculty of Mathematics and Physics, University of Ljubljana, Jadranska 19, SI-1000 Ljubljana, Slovenia}

\begin{abstract}
Very high-energy astrophysical gamma rays suffer a suppression of their flux along their propagation due to their interaction, through the $\gamma\gamma\to e^+e^-$ pair-production process, with the soft photon backgrounds present in the Universe. We examine the Universe's transparency to gamma rays within a Lorentz Invariance Violation (LIV) framework, focusing on photon subluminal quadratic corrections driven by a high-energy scale. Based on an explicit calculation, we provide a new expression for the cross section that overcomes the limitations of previous approaches and refines existing constraints for the LIV scale, while we introduce a new approximation that may be useful in LIV scenarios beyond effective field theory. These improvements appear essential for setting constraints on LIV effects with future observations at ultra-high energies, where previous approximations may fall short.
\end{abstract}

\maketitle

\section{Introduction}
\label{introduction}

The advancement of gamma-ray astronomy in the last decades, allowing detection of high energy (HE, 100\,MeV -- 100\,GeV), very-high energy (VHE, 100\,GeV -- 100\,TeV), and ultra-high energy (UHE, 100\,TeV -- 100\,PeV) gamma rays has given us a realistic opportunity to test fundamental physics. 
This is the case of quantum gravity (QG) phenomenological models that consider a violation of the space-time symmetries of special relativity (SR), which could be manifest at high enough energies~\cite{Addazi:2021xuf}. The combination of high energies and astrophysical distances, which serve as an amplifier, makes gamma-ray astronomy a solid ground to test effects of Lorentz invariance violation (LIV)~\cite{Amelino-Camelia:1997ieq,Mattingly:2005re,Liberati:2013xla}. 

LIV corrections to SR are usually parameterized by high-energy (`quantum gravity') scales $E_{\text{LIV},n}$. These scales are introduced at the Lagrangian level in the effective field theory framework, suppressing the higher dimensional LIV operators. Generally speaking, $E_{\text{LIV},n}$ can have different values for different particles. In the case of the photon, they lead to a modified photon dispersion relation of the general form
\begin{equation}
E^2 - \vec{k}^{\,2} = E^2 \sum_{n=1}^{\infty}S_n \left(\frac{E}{E_{\text{LIV},n}}\right)^n\,,
\label{eq:mdr}
\end{equation}
where $E$ and $\vec{k}$ are the energy and momentum of the photon, respectively. This is the usual starting point of experimental tests of LIV. 
Bounds on  the different scales $E_{\text{LIV},n}$ can be put from data at much lower energies, so that $E\ll E_{\mathrm{LIV,}n}$, and usually only linear ($n=1$) or quadratic ($n=2$) corrections are investigated independently, assuming they are the  dominant terms in Eq.~(\ref{eq:mdr}), respectively.
Considering each term in the series separately, the parameters $S_n = \pm 1$ allow for the possibility of superluminal (photon velocity larger than the standard speed of light) and subluminal (photon velocity smaller than the standard speed of light) behaviour.
The consequences of a modified dispersion relation include an energy-dependent photon group velocity, which is tested by searching for time delays in the detection of photons of different energies (see, e.g.~\cite{Ellis:2002in,Ellis:2005sjy,Martinez:2008ki, Albert:2007qk, Vasileiou:2013vra, Wei:2017qfz, Ellis:2018lca, MAGIC:2020egb, Bolmont:2022yad, Abe_2024}); birefringence effects, where the photon group velocity depends on polarisation, as well as energy (see, e.g.,~\cite{Kostelecky:2008be, Kislat:2017kbh, Friedman:2020bxa, Toma:2012xa, Kostelecky:2013rv, Gotz:2014vza}); photon instability (see, e.g.,~\cite{HAWC:2019gui, LHAASO:2021opi}); and anomalous electromagnetic interactions, which can be manifested as changes in the opacity of the Universe to gamma rays, synchrotron emission, Compton scattering, extensive air shower development, etc. (see, e.g.,~\cite{Biteau:2015xpa, Abdalla:2018sxi, Lang:2018yog, Jacobson:2002ye, Abdalla:2018sxi, Rubtsov:2016bea}). See~\cite{Terzic:2021rlx} for a comparison between different tests performed on gamma rays.

Experimental constraints on the new physics depend on the specific effect and the LIV scenario under consideration\footnote{For an up to date census of experimental tests and constraints on the quantum-gravity scale, see the QG-MM Catalogue~\cite{QGMMCatalogue}.}.  
For linear corrections, birefringence effects establish the lower bound on $E_{\text{LIV},1}$ to be many orders of magnitude above the Planck scale~\cite{Gotz:2014vza, Whittaker:2017hnz}. They are, however, absent in the quadratic case. In addition, superluminal scenarios are very much constrained by the absence of vacuum pair production or photon splitting~\cite{LHAASO:2021opi}. In the $n=1$ case, the bounds on $E_{\text{LIV},1}$ surpass again the Planck scale by several orders of magnitude, while in the $n=2$ case, present bounds on $E_{\text{LIV},2}$  reach three orders of magnitude below the Planck energy. 

Constraints obtained from time delays are usually weaker than the ones obtained from other effects. The time delay bounds on the linear correction are of the order of the Planck energy~\cite{Vasileiou:2013vra}, several orders of magnitude below the bounds coming from birefringence. The time delay bounds on the quadratic correction are eight orders of magnitude below the Planck scale~\cite{Abdalla:2019krx}, which are approximately of the same order as bounds obtained from anomalous photon interaction effects in the subluminal case~\cite{Abdalla:2019krx}.
This makes the subluminal quadratic correction scenario an especially interesting one, with two complementary phenomenological windows. Moreover, studies on the LIV modification interactions typically rely on certain approximations in the interaction cross section, which may affect the inferred constraints. Our goal in this paper is to critically examine the procedures employed to search for the effects of modified interactions of high-energy photons with the electromagnetic backgrounds, namely the Cosmic Microwave Background (CMB) and the Extragalactic Background Light (EBL, optical and infrared photons), in this particular LIV scenario.

Previous phenomenological approaches and experimental tests of the Universe's transparency to gamma rays considered LIV modifications to the pair-creation process $\gamma\gamma\to e^+e^-$ only at the level of a modification in the threshold of the reaction, disregarding any changes in its cross section~\cite{Martinez-Huerta:2020cut,Blanch:2001hu}, or assumed an ad-hoc modification of the cross section, simply by replacing the expression of the square of the total momentum of the two photons in SR by the new expression in the LIV case~\cite{Tavecchio:2015rfa,Abdalla:2018sxi,Fairbairn:2014kda,Abdalla:2019krx}. As we will see in Sec.~\ref{sec:interaction}, both of these approaches produce unphysical distortions close to the threshold of the reaction. 

In most of the cases, modifications in the fermion sector were disregarded following experimental constraints more stringent than for the photon sector~\cite{Maccione:2008iw, Galaverni:2007tq,Stecker:2001vb, Li:2022ugz, Maccione:2007yc,Liberati:2012jf, Rubtsov:2016bea}. For instance, the absence of vacuum Cherenkov radiation from Crab Nebula observations, put the constraints in the superluminal case to be at least six orders of magnitude above the Planck scale and three orders of magnitude below the Planck scale for the linear and quadratic cases, respectively~\cite{Li:2022ugz}. Similarly, for the case of subluminal electrons, the scale of new physics in the linear case is constrained to be at least five orders of magnitude above the Planck scale, while in the quadratic case it is three orders of magnitude below the Planck scale~\cite{Li:2022ugz, Rubtsov:2016bea}, higher 
than the scales under study in the physics of photon anomalous interactions.

To our knowledge, currently the only derivation of the pair-production cross section by introducing LIV operators at the Lagrangian level has been done in \cite{Rubtsov:2012kb}. There, the influence of dimension-five operators, corresponding to $n=1$, were considered negligible because of birefringence constraints on the linear correction. Operators introducing LIV in both photon and fermion sectors were considered, which rendered the calculation rather cumbersome, forcing the authors to perform it in the leading-log approximation. This approximation is not valid near the reaction thresholds. Moreover, no study on the impact of such limitation in the computation of the photon mean free path was performed.

The aim of this work is to compare previous approximations and propose new approaches for studying the effects of quadratic ($n=2$) corrections to the photon dispersion relation on the transparency of the Universe to gamma rays in the subluminal scenario. In Section~\ref{sec:interaction}, we present threshold conditions, effective approximations to the cross section that capture the influence of the modified kinematics, and results from explicit calculations of the cross section that include LIV effects. In Section~\ref{sec:transp}, these approaches are compared based on their predictions for the mean free path and survival probability.  We lay out our conclusions in Section~\ref{sec:conclusion}. 

\section{Photon-photon interaction}
\label{sec:interaction}

The dominant interaction responsible for the absorption of VHE and UHE gamma rays by the electromagnetic background is the production of electron-positron pairs~\cite{Nikishov:1962,Gould:1967zza,Gould:1967zzb}. Let us consider the emission of an electron and positron with energy-momentum $p_-=(E_-,\vec{p}_-)$ and $p_+=(E_+,\vec{p}_+)$,  by the electromagnetic interaction of a high-energy photon, of energy-momentum $k=(E,\vec{k})$, with a low-energy background photon, of energy-momentum $q=(\omega,\vec{q})$,
\begin{equation}\label{pairc}
            \gamma(k) + \gamma_\text{soft}(q) \to e^-(p_-) + e^+(p_+).
\end{equation}
The threshold condition for this reaction to occur can be obtained by imposing the conservation of the squared total four-momentum of the system before and after the scattering,
\begin{equation}
 (k+q)^2 \,=\, (p_-+p_+)^2 \geq 4 m_e^2\,.
 \label{eq:genthreshold}
\end{equation}
The cross section of the process must contain information about the normalization of the initial states, and an integral of the squared quantum mechanical matrix amplitude of the process. This can be written as
\begin{equation}
    \sigma=\frac{1}{\mathcal{K}}\, \times \mathcal{F}\,, \quad\text{with}\quad \mathcal{F}=\int [\mathcal{DPS}] \, |\mathcal{M}|^2, 
    \label{eq:cross}
\end{equation}
where we have introduced the notation $[\mathcal{DPS}]$ for the integral measure over the final particle phase space, $\mathcal{M}$ for the matrix element, and $1/\mathcal{K}$ for the initial state factor.
For the photon-photon pair production, $\mathcal{F}$ is given by
\begin{align} 
    &\mathcal{F}(E, \omega, \theta) = \int \frac{d^3 \vec{p}_-}{(2\pi)^3 2 E_-} \, \frac{d^3 \vec{p}_+}{(2\pi)^3 2 E_+} \; |\mathcal{M}_{\gamma\gamma\rightarrow e^-e^+}|^2 \notag \\ & \times (2\pi)^4 \delta(E+\omega-E_--E_+)\, \delta^3(\vec{k}+\vec{q}-\vec{p}_--\vec{p}_+) \,,
\end{align}
where $\theta$ is the angle between the directions of the two photon momenta $\vec{k}$ and $\vec{q}$.

\subsection{Special relativity}

Particularizing the previous discussion to the case of SR, the threshold condition (Eq.~\eqref{eq:genthreshold}) takes the form
\begin{equation}
    2E\omega(1-\cos\theta) \geq 4m_e^2\,,
    \label{BW_threshold}
\end{equation}
which encourages us to define a new variable, 

\begin{equation}\label{eq:s_bar}
\bar s \doteq \frac{ 2E\omega(1-\cos\theta)}{4m_e^2}\,.
\end{equation}
The threshold condition can now be stated as $\bar s \geq 1$\footnote{In the whole manuscript, we use barred variables for dimensionless quantities.}.
Equivalently, the SR cross section $\sigma_\text{SR}$ is given by the product of the initial state factor, $1/\mathcal{K}_\text{SR}(E, \omega, \theta)= 1/4 E\omega (1-\cos\theta)$, and the corresponding integral of the amplitude over the final phase space, $\mathcal{F}_\text{SR}(E,\omega,\theta)$. Let us note that using Eq.~\eqref{eq:s_bar}, the inverse of the SR initial state factor can be written as $\mathcal{K}_\text{SR}=8m_e^2 \bar s$, which is only function of $\bar s$. Similarly, the result of the integral over the phase space can be written as

\begin{align}
    {\cal F}_\text{SR}(E,\omega,\theta) = 4\pi \alpha^2 \Bigg[ &\left(2 + \frac{2}{\bar s}-\frac{1}{\bar{s\,}^2} \right) \ln \left(\frac{1+\sqrt{1-1/\bar s}}{1-\sqrt{1-1/\bar s}}\right) \notag \\ &- \left(2+\frac{2}{\bar s}\right) \sqrt{1-1/\bar s} \Bigg] \doteq \mathcal{F}_\text{BW}(\bar s) \,.
    \label{eq:BW_F}
\end{align}
This expression is only defined for $\bar s \geq 1$ and is the well-known result usually referred to as the Breit-Wheeler formula~\cite{Breit:1934zz}.

\subsection{Lorentz invariance violation}

From the perspective of effective field theory, one could perform a systematic analysis (see, e.g.,~\cite{Colladay:1998fq, mattingly2008tested, Mattingly:2005re}) of all the LIV operators that could be added to the QED Lagrangian, subject to additional restrictions, such as gauge and rotational invariance. As explained in the Introduction, we do not consider LIV terms originating from dimension-five operators in the photon sector. Furthermore, we neglect all LIV terms involving the fermion field, since their coefficients are subject to strong phenomenological or experimental constraints. There are still several terms available that are quadratic in the gauge field, which prove to be equivalent for our purposes and provide us with the same modified dispersion relation,
\begin{equation}
E^2 - \vec{k}^2 \,=\, -\frac{E^4}{\Lambda^2}\,,
\label{eq:mdr2}
\end{equation}
which is simply the $n=2$ case of Eq.~\eqref{eq:mdr} for subluminal photons. From now on, for shortness, we substitute $E_{\text{LIV},2}$ with $\Lambda$.

We aim to investigate the observable effects resulting from Eq.~\eqref{eq:mdr2} on the propagation of gamma rays across the Universe, i.e., in the transparency of the Universe. Naively, one would expect that the quadratic modification in Eq.~\eqref{eq:mdr2} would produce corrections of order $(E/\Lambda)^2$ with respect to the result of SR. Given that the maximum detected energy of gamma rays is of the order of the petaelectronvolt (PeV)~\cite{LHAASO:2021gok}, and that the lower bounds for $\Lambda$ from time-of-flight analyses in the quadratic case are of the order $10^{-8}\,E_\text{Pl}$~\cite{Addazi:2021xuf} (with $E_\text{Pl}\approx 1.22 \times 10^{13}\,\mathrm{PeV}$ the Planck energy), we can write
\begin{equation}
    (E/\Lambda)^2 \approx 10^{-10} (E/\text{PeV})^2 (10^{-8}\,E_\text{Pl}/\Lambda)^2 \,,
\end{equation}
so that these corrections would be totally unobservable.
It is easy to check that this naive expectation is not correct from the following simple kinematical considerations. Due to the modified energy-momentum relation of the gamma ray, the threshold condition now reads
\begin{equation}
    2E\omega(1-\cos\theta) - E^4/\Lambda^2 \geq 4m_e^2\,,
    \label{eq:threshold}
\end{equation}
where the left-hand side of the inequality is the value of the total four-momentum before the scattering, $(k+q)^2$, modified from its SR value by LIV, and the right-hand side is the minimum value of the total four-momentum after the scattering, $(p_-+p_+)^2$, which is $4 m_e^2$ as in SR, since the dispersion relation of the electron and the positron are not modified.
This encourages us to define a new variable,
\begin{equation}
    \bar{\tau}\doteq\bar{s}-\bar{\mu}, \quad \text{where } \bar{\mu}\doteq\frac{E^4}{4m_e^2\Lambda^2} \,,
    \label{taubar}
\end{equation}
such that the threshold condition can be stated as $\bar \tau \geq 1$. The variable $\bar{\mu}$  controls how much the result of LIV differs from the case of SR. We see then that the effect of LIV in the threshold condition can be observable even when $E\ll \Lambda$. 

Depending on the value of the scale of new physics, the quartic equation on the energy of the gamma ray, $E$, corresponding to the equality in Eq.~\eqref{eq:threshold}, will either have two (positive real) solutions, $E_\text{th}^{(1)} \leq E_\text{th}^{(2)}$, which will restrict the values of $E$ to the range $E_\text{th}^{(1)} \leq E \leq  E_\text{th}^{(2)}$; or will not have any real solution, completely suppressing the photon-photon interaction.
\begin{figure}[t]
    \centering \includegraphics[width=\linewidth]{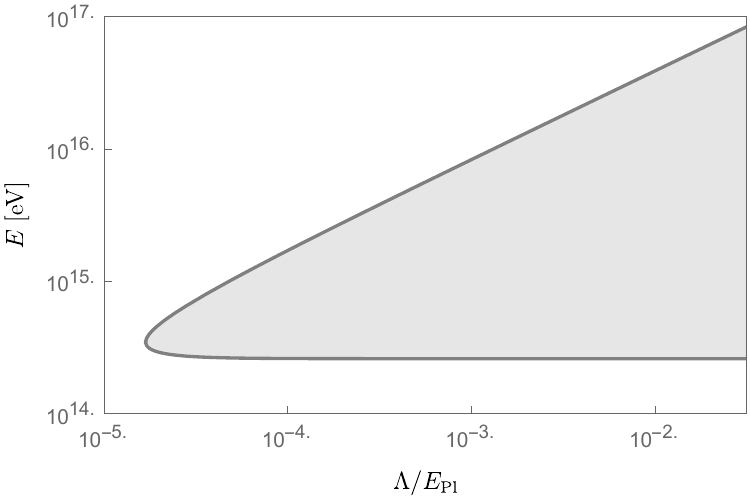}
    \caption{Kinematically allowed region (grey area) for the energy of the high-energy photon $E$ and the scale of new physics $\Lambda$ for the process  $\gamma+ \gamma_\text{soft} \to e^- + e^+$, for a subluminal quadratic correction in the dispersion relation, for a fixed low-energy photon energy $\omega=10^{-3} \mathrm{eV}$, and for a fixed angle $\theta=\pi$. The dark grey line represents the set of points that satisfy the threshold equality in Eq.~\eqref{eq:threshold}.}
    \label{fig:Eth}
\end{figure}
The analytical expressions for the two solutions do not have a simple form. However, if one writes Eq.~\eqref{eq:threshold} in terms of $\bar s$ (for $\omega$ and $\theta$ fixed), the quartic equation of $E$ can be written as
\begin{equation}\label{eq:sbarsol}
    1 - \bar{s} + a \,\bar{s}^4 \,=\, 0\,, \quad\text{with}\quad a=\frac{4m_e^6}{\omega^4(1-\cos\theta)^4\Lambda^2} \,,
\end{equation}
which has two positive real solutions for $\bar s$ when $a<27/256$. Under the assumption $a \ll 1$, if one rewrites the threshold condition as $a=(\bar{s}-1)/\bar{s}^4$, the two solutions meeting the approximation are easily seen to be
\begin{equation}
    \bar{s}^{(1)} \approx 1\,, \quad\text{and}\quad  \bar{s}^{(2)} \approx a^{-1/3} \,.
\end{equation}
Solving for $E$, one obtains that the values of the corresponding thresholds approximately are 
\begin{equation}
E_\text{th}^{(1)} \approx \frac{2 m_e^2}{\omega(1-\cos\theta)}\,, \quad\text{and}\quad E_\text{th}^{(2)} \approx \left[2 \omega (1-\cos\theta) \Lambda^2\right]^{1/3}\,, 
\end{equation}
in agreement with the allowed region in Fig. \ref{fig:Eth}, in which one of the solutions swiftly converges to the solution of SR and the other one grows with $\Lambda$ (going to infinity in the SR limit).

Besides the modification in the threshold condition, LIV corrections are also expected to change the cross section of the photon-photon pair production process. Recalling the decomposition previously mentioned in Eq.~\eqref{eq:cross}, one could write the modified cross section as
\begin{equation}
    \sigma_\text{LIV}(E, \omega, \theta) = \frac{1}{\mathcal{K}_\text{LIV}(E, \omega, \theta)} \; {\cal F}_\text{LIV}\left(E, \omega, \theta\right)\,,
    \label{eq:LIV_cross_sec}
\end{equation}
with $1/\mathcal{K}_\text{LIV}$ the modified kinematical factor and $\mathcal{F}_\text{LIV}$ the new result of the integral of the matrix element over the two final particle phase space, which is only defined when the threshold condition, Eq.~\eqref{eq:threshold}, is satisfied. However, the modification due to LIV in the initial state kinematical factor can only come from the modification of the dispersion relation of the high-energy photon, and, consequently, it can only produce a negligible correction of order $(E/\Lambda)^2$. Taking that into account, it is safe to assume that $\mathcal{K}_\text{LIV}\approx\mathcal{K}_\text{SR}= 4E\omega(1-\cos\theta)$ and, as a consequence, all the observable effects of LIV in the cross section are contained in the function $\mathcal{F}_\text{LIV}$.

\subsection{Effective approaches}
\label{sec:efective}

Given the difficulty to compute the integral of the matrix element over the two particle phase space, some simple approximations have been proposed in the literature for the modified cross section of the photon-photon pair production. These proposals can be interpreted as approximations of the function $\mathcal{F}_\text{LIV}$ (following the previously discussed argument that one can disregard the effects of LIV in the kinematical prefactor $1/\mathcal{K}_\text{LIV}$).

The simplest proposal is to consider that the new cross section is still the Breit-Wheeler one, but it now takes non-zero values only when the modified threshold condition, Eq.~\eqref{eq:threshold}, is satisfied~\cite{Martinez-Huerta:2020cut, Blanch:2001hu}. This proposal corresponds to approximating the function $\mathcal{F}_\text{LIV}$ by
\begin{equation}\label{approx1}
    \mathcal{F}_\text{LIV}(E, \omega, \theta) \approx \mathcal{F}_\text{BW}\left(\bar s \right) \doteq \mathcal{F}_\text{LIV}^{(1)}(\bar\tau,\bar\mu)\,,
\end{equation}
for $\bar\tau\geq 1$, or equivalently for $\bar s\geq 1 + E^4/(4m_e^2\Lambda^2)$, and zero otherwise. This corresponds to the black curves in Fig.~\ref{fig:comparison}, where one can see that, since the function $\mathcal{F}_\text{BW}$ does not go to zero when $E$ approaches the second threshold, a discontinuity is generated. 

The second proposal mentioned in the literature \cite{Tavecchio:2015rfa,Abdalla:2018sxi,Fairbairn:2014kda,Abdalla:2019krx} tries to improve the discontinuity produced by the previous approximation. Taking into account that $\bar\tau$ goes to 1 as we approach the second threshold, and that also $\sigma_\text{SR}(1)=0$, this method proposes the replacement of $\bar s$ by $\bar \tau$ in the SR cross section. However, doing this replacement at the level of the cross section not only modifies the function $\mathcal{F}$, but also introduces an additional non-negligible correction in the kinematical prefactor $1/\mathcal{K}$, which is no longer of order $(E/\Lambda)^2$, and, consequently, introduces an unjustified extra correction in the cross section. An equivalent way to write this modified cross section is to absorb the extra correction inside the function $\mathcal{F}$ (instead of inside the kinematical prefactor $1/\mathcal{K}$), so that $\mathcal{F}_\text{LIV}$ becomes
\begin{equation}\label{approx2}
    \mathcal{F}_\text{LIV}(E, \omega, \theta) \approx \left(1+\frac{\bar \mu}{\bar \tau}\right) \, \mathcal{F}_\text{BW}(\bar \tau) \doteq  \mathcal{F}_\text{LIV}^{(2)} (\bar\tau,\bar\mu) \,,
\end{equation}
which is only defined when $\bar\tau\geq 1$, i.e., when the threshold condition, Eq.~\eqref{eq:threshold}, is satisfied. This approximation corresponds to the blue curve in Fig.~\ref{fig:comparison}, which differs significantly from the other approaches close to the second threshold. This is because the quotient $(\bar \mu/\bar\tau)$ is much larger than 1 for $E \lesssim E_\text{th}^{(2)}$.

If one does the replacement of $\bar s$ by $\bar \tau$ only inside the function $\mathcal{F}$, instead of doing it at the level of the cross section (i.e., disregarding any modification in the kinematical factor $1/\mathcal{K}$), one obtains a third proposal which still fulfills the desired smooth behaviour at the second threshold but without the anomalous extra correction,
\begin{equation}
\mathcal{F}_\text{LIV}(E, \omega, \theta) \approx \mathcal{F}_\text{BW}(\bar \tau) \doteq \mathcal{F}_\text{LIV}^{(3)}(\bar\tau, \bar\mu)\,,
\label{approx3}
\end{equation}
which again is only defined when $\bar\tau\geq 1$, i.e., when the threshold condition, Eq.~\eqref{eq:threshold}, is satisfied. This approximation corresponds to the green curves in Fig.~\ref{fig:comparison}.

\subsection{Explicit calculations}
\label{sec:explicit}

As discussed before, an explicit computation of the integral of the squared matrix element over the two fermion phase space is rather difficult. However, in the literature there have been some attempts to calculate the function $\mathcal{F}_\text{LIV}$ directly from a Lagrangian, under certain approximations.

\begin{figure}[t]
    \centering
    \includegraphics[width=\linewidth]{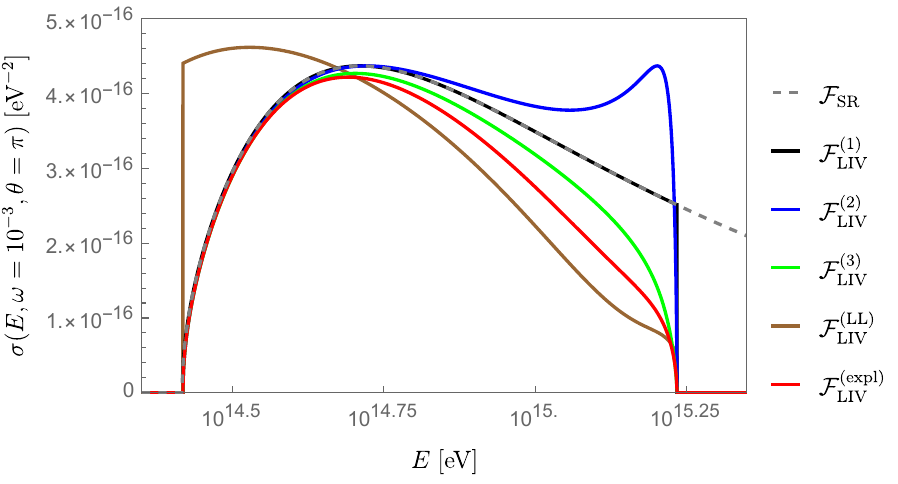}
    \includegraphics[width=\linewidth]{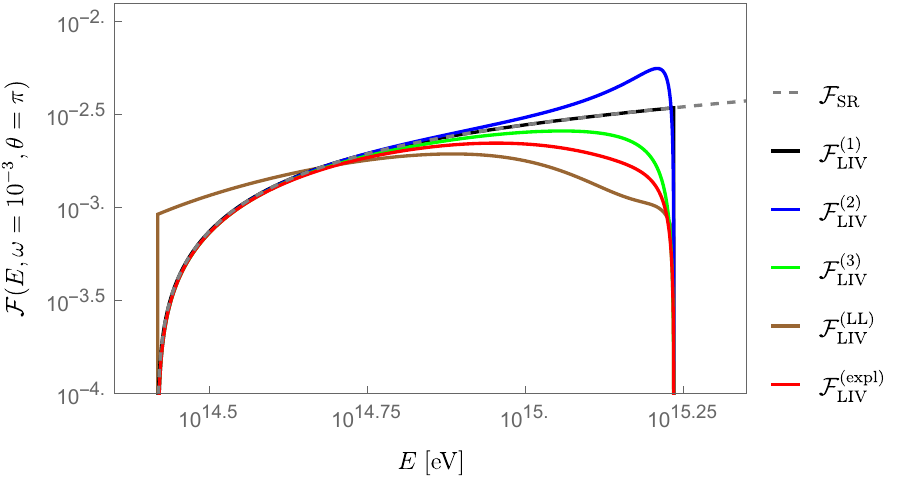}
    \caption{Comparison of the cross section $\sigma_\text{LIV}$ (upper plot) and the function $\mathcal{F}_\text{LIV}$ (bottom plot) for the explicit (red) and leading-log computations (brown) discussed in Section~\ref{sec:explicit}, and the approximations discussed in Section~\ref{sec:efective} (black, blue and green). We consider a fixed energy of the low-energy photon, $\omega=10^{-3}\,\mathrm{eV}$, a fixed angle, $\theta=\pi$, and a value of the scale of new physics $\Lambda/E_\text{Pl}=10^{-4}$. The grey dashed line represents the SR (Breit-Wheeler) case.}
    \label{fig:comparison}
\end{figure}

In~\cite{Rubtsov:2012kb}, a calculation was carried out for initial state energies far above the threshold requirements, which gives us the so-called leading-log (LL) result. In addition to the modified dispersion relation for the photon, analogous modifications were also introduced in the fermionic sector. Furthermore, the massless limit for the fermions was employed from the beginning\footnote{The mass was ultimately reintroduced in order to avoid the logarithmic divergence of the final state phase space integral.}. In order to be able to make their result comparable with a calculation which does not include fermionic corrections, we set them to zero and rewrite it in terms of the variables $\Bar{\tau}$ and $\Bar{\mu}$,
\begin{equation}
    \mathcal{F}_\text{LIV}^\text{(LL)}(\bar\tau,\bar\mu) \doteq  4\pi \alpha^2 \left(1+\frac{(\bar\tau-\bar\mu)^2}{(\bar\tau+\bar\mu)^2}\right) \ln(\bar\tau)
    \,.
    \label{eq:FLL}
\end{equation}
This will be a good approximation as long as one is far from the reaction thresholds, i.e., $\bar \tau \gg 1$. However, the study of the photon absorption in the electromagnetic background requires to go beyond this limit, which fails at the thresholds ($\bar \tau = 1$), as can be seen in Fig.~\ref{fig:comparison}.

An alternative calculation, which does not rely on the $\bar{\tau}\gg1$ limit, has been performed in~\cite{Vrban:2022}. This was feasible because LIV fermionic corrections were neglected, which also allowed for the fermionic mass to be kept from the beginning. 
The strategy followed in the computation consists on splitting the sum over the two photon polarizations ($\xi=1,2$) into two terms, one that coincides with the usual sum of SR, and a correction term\footnote{Gauge invariance implies the relation $k^\mu \mathcal{M}_\mu=0$. In SR this leads to the relation $|\mathcal{M}_0|^2 = |\mathcal{M}_3|^2$, which justifies the covariance of the sum over polarizations. The modification of the relation between $|\mathcal{M}_0|^2$ and $|\mathcal{M}_3|^2$, due to the modification of the dispersion relation in (\ref{eq:mdr}), is the origin of the second term in the sum over polarizations in Eq.~\eqref{pol-sum}.} proportional to $(E/\Lambda)^2$, i.e.,

\begin{align}
    \sum_{\xi=1,2} \varepsilon^\mu(\vec{k}, \xi) &\varepsilon^{\nu *}(\vec{k}, \xi) \mathcal{M}_\mu \mathcal{M}_\nu^* \notag \\
    \approx & - g^{\mu\nu} \mathcal{M}_\mu \mathcal{M}_\nu^* + \left(E/\Lambda\right)^2 \mathcal{M}_0 \mathcal{M}_0^*\,.
    \label{pol-sum}
\end{align}
The first term produces a squared matrix element which coincides with the SR result when expressed in terms of the momenta ($q, p_-, p_+$) of the low-energy photon and the two fermions, which does not contain any effect of LIV. In contrast, the correction term will produce an additional squared matrix element which comes purely as a consequence of LIV. After integrating the sum of both matrix elements over the modified two fermion phase space, one obtains an explicit formula for $\mathcal{F}_\text{LIV}$. When the hierarchy of scales $\Lambda \gg E\gg m_e\gg \omega$ is employed, it can be written in the following compact form

\begin{align}
\mathcal{F}_\text{LIV}^\text{(expl)}(\bar\tau,\bar\mu) \doteq\, 4 \pi \alpha^2 \Biggl[&\left(2+\frac{2\bar\tau(1-2\bar\mu)}{(\bar\tau+\bar\mu)^2}-\frac{(1-\bar\mu)}{(\bar\tau+\bar\mu)^2}\right) \notag \\ \times \ln\left(\frac{1+\sqrt{1-1/\bar\tau}}{1-\sqrt{1-1/\bar\tau}}\right)
- &\left(2+\frac{2\bar\tau(1-4\bar\mu)}{(\bar\tau+\bar\mu)^2}\right) \sqrt{1-1/\bar\tau} \Biggr]\,,
\label{eq:F_complete}
\end{align}
which reduces to the SR result in the limit $\Lambda\to\infty$ (or equivalently $\bar\mu\to 0$). Analogously, taking the $\Bar{\tau}\gg 1$ limit reproduces the leading-log result, Eq.~\eqref{eq:FLL}.

Eq.~\eqref{eq:F_complete} provides, up to now, the most complete calculation of the function $\mathcal{F}_\text{LIV}$ for the subluminal quadratic case, and so of the corresponding LIV photon-photon pair production cross section. Additionally, Eq.~\eqref{eq:F_complete} also allows us to compare the adequacy of the effective approximations studied in Section~\ref{sec:efective}. The comparison is shown in the plots of Fig.~\ref{fig:comparison}, where the explicit result is shown in red. The approximation in Eq.~\eqref{approx3} (green curves) is the effective proposal which gets the closest to the explicit calculation behaviour. In contrast, the approximation in Eq.~\eqref{approx2} (blue curves), due to the artificially introduced extra correction, differs from the expected behaviour and, as we will see in the next section, can produce an overestimation of the absorption effect, or equivalently, an underestimation of the bounds on the scale of new physics $\Lambda$. 
The first approximation (black curves, Eq.~\eqref{approx1}) involves a discontinuity, while the second one (blue curves, Eq.~\eqref{approx2}), develops a sharp peak towards the upper reaction threshold. We consider both behaviours unphysical, and therefore discourage the use of these approximations in favour of the explicit result (Eq.~\eqref{eq:F_complete}).

For the linear case, which was not considered in this work, birefringence constraints call for going beyond effective field theory, which forces us to consider effective approaches. However, it seems advisable to apply the analog of the third approximation Eq.~\eqref{approx3} (green) to this case, rather than the analogs of the other two effective approaches from Eq.~\eqref{approx1} and Eq.~\eqref{approx2} (black and blue), for the same reasons discussed for the quadratic case.

\section{Universe transparency}
\label{sec:transp}

Transparency of the Universe to gamma rays is usually characterized by the opacity $\tau(E,z_s)$, which is related with the photon survival probability according to
\begin{equation}\label{eq:prob}
    \mathrm{Prob}(E,z_s) = \exp\left(-\tau(E,z_s)\right) \,,
\end{equation}
where $E$ is the observed gamma-ray energy and $z_s$ is the redshift of the observed source. This fundamental observable can be computed using the cross section of the dominant photon absorption process, which carries the information of the probability of a single interaction, multiplied by the spectral density of background photons, and integrated to all the angles and over all the trajectory to the detector,

\begin{align}\label{eq:opa}
    \tau(E,z_s) \,=\, &\int_{0}^{z_s}dz\,\frac{dl}{dz}\int_{-1}^{1} d\cos\theta \left(\frac{1-\cos\theta}{2}\right) \notag \\ \times &\int_{\omega_\text{th}(E,\theta)}^\infty d\omega \; n(\omega,z) \,\sigma(E(1+z),\omega,\theta) \,.
\end{align}
$n$ represents the spectral density of the low-energy electromagnetic background, 
and $\omega_\text{th}$ is the minimum value of the energy of the soft photon necessary to satisfy the threshold condition.

\subsection{Galactic sources}

Here we focus on galactic sources, for which one can neglect the redshift dependencies in Eq.~\eqref{eq:opa}. Consequently, the first integral in Eq.~\eqref{eq:opa} is just the Euclidean distance to the source $d_s = \int_0^{z_s} (dl/dz)\, dz$, and the background photon density is constant with respect to the distance from the Earth.
Dividing Eq.~\eqref{eq:opa} by $d_s$, one can recognize the inverse of the mean free path
\begin{equation}
    \frac{1}{\lambda(E)} = \int_{-1}^{1} d\cos\theta \left(\frac{1-\cos\theta}{2}\right) \int_{\omega_\text{th}(E,\theta)}^\infty \hspace{-1.5em} d\omega \; n(\omega) \,\sigma(E,\omega,\theta) \,.
\end{equation}
We will write the probability of survival of gamma rays travelling from a distance $d_s$ as
\begin{equation}
    \mathrm{Prob}(E,d_s) \approx \exp\left(-d_s/\lambda(E)\right) \,.
\end{equation}

At these close distances, the gamma-ray absorption on the EBL is rather small, making the possible effects of LIV virtually negligible. The CMB has the photon density more than two orders of magnitude higher, providing considerably more targets for gamma-ray scattering. This, however, requires higher gamma-ray energies in order for the reaction threshold to be reached. 
The CMB spectral density has a well known analytical form given by the black body emitted spectrum,
\begin{equation}\label{eq:ncmb}
    n_{\text{CMB}}(\omega) = \frac{(\omega/\pi)^2}{\exp(\omega/(k_\text{B} T_0))-1} \,,
\end{equation}
where $T_0=2.73 \,\mathrm{K}$ at present, and $k_\text{B}$ is the Boltzmann constant.

It proves useful to make a change of variables from $(\theta,\omega)$ to the dimensionless variables $(\bar\tau,\bar\omega)$, where we defined $\bar\omega=\omega/(k_\text{B}T_0)$ for the CMB. Then, the inverse of the mean free path can be written as

\begin{align}\label{eq:mfp}
    \frac{1}{\lambda(E)}= \frac{m_e^2k_\text{B}T_0}{4\pi^2 E^2} &\int_1^\infty d\bar\tau \, \mathcal{F} (\bar\tau,\bar\mu)  \notag \\ \times &\int_{\bar\omega_\text{th}(\bar\tau,\bar\mu)}^{\infty}d\bar\omega \; \frac{1}{\exp{(\bar\omega)}-1}\,,
\end{align}
where now
\begin{equation}
    \bar\omega_{\text{th}}(\bar\tau,\bar\mu)=\frac{m_e^2}{k_{\text{B}}T_0E}\;(\bar \tau + \bar\mu)\,,\ \text{with } \ \bar\mu=\frac{E^4}{4 m_e^2 \Lambda^2}.
\end{equation}
By $\mathcal{F}(\bar\tau,\bar\mu)$ in Eq.~\eqref{eq:mfp} we mean any of the approaches mentioned in Section~\ref{sec:efective} and Section~\ref{sec:explicit}, or the SR case which corresponds to taking $\bar\mu\rightarrow 0$ (which also implies $\bar\tau \rightarrow \bar s$). Let us note that, as expected, observables like the opacity do not depend on the kinematical prefactor $1/\mathcal{K}$ in the cross section, but only on the function $\mathcal{F}$.

\begin{figure}[p]
    \centering
    \includegraphics[width=\linewidth]{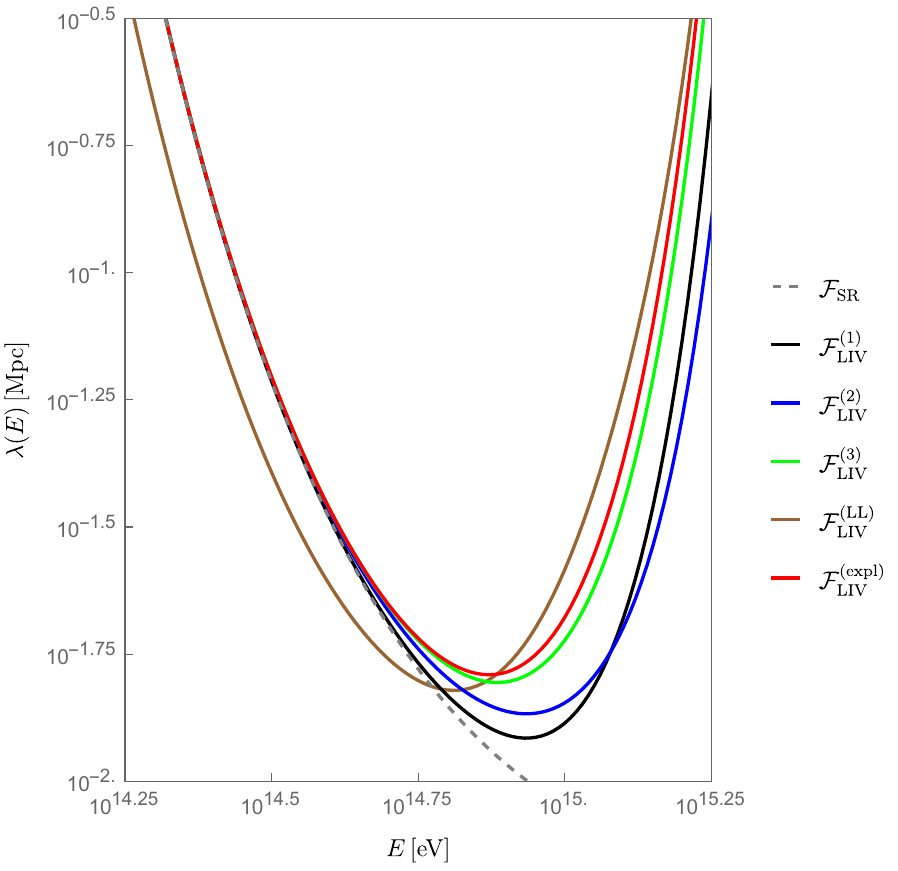}
    \caption{Comparison of the mean free path for the explicit (red) and leading-log computations (brown) discussed in Section~\ref{sec:explicit}, and the approximations (black, blue and green) discussed in Section~\ref{sec:efective}. We consider interactions with the CMB, and a value of the scale of new physics $\Lambda/E_\text{Pl}=10^{-4}$.}
    \label{fig:comparison_path}
\end{figure}

\begin{figure}[p]
    \centering
    \includegraphics[width=\linewidth]{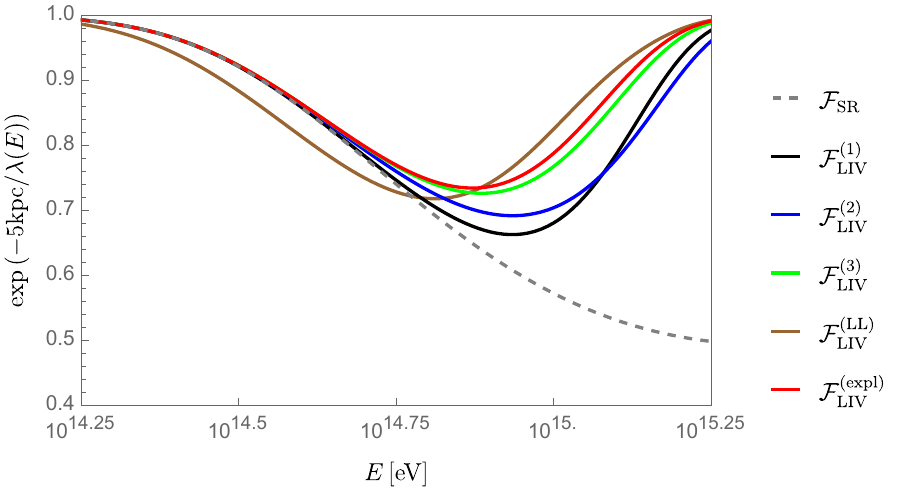}
    \includegraphics[width=\linewidth]{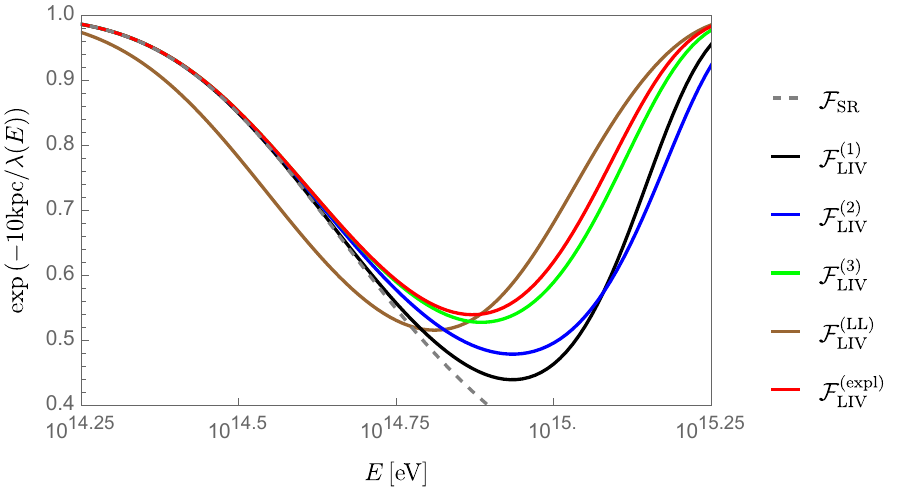}
    \caption{Comparison of the survival probability for the explicit (red) and leading-log computations (brown) discussed in Section~\ref{sec:explicit}, and the approximations (black, blue and green) discussed in Section~\ref{sec:efective}. We consider interactions with the CMB, a value of $\Lambda/E_\text{Pl}=10^{-4}$, and two source distances, $d_s=5\mathrm{kpc}$ (upper plot) and $10\mathrm{kpc}$ (bottom plot).}
    \label{fig:comparison_prob}
\end{figure}

In Figs.~\ref{fig:comparison_path} and~\ref{fig:comparison_prob}, we show a comparison of the mean free path and survival probability, respectively, using the three different approximations introduced in Section~\ref{sec:efective}, Eqs.~\eqref{approx1}, \eqref{approx2} and \eqref{approx3}; and the leading-log and explicit results of Section~\ref{sec:explicit}, Eqs.~\eqref{eq:FLL} and \eqref{eq:F_complete}. One can check that, as we anticipated from the behaviour of the cross sections in Fig.~\ref{fig:comparison}, the use of the first (black) and second (blue) approximations introduces an overestimation of the absorption effect, which leads to underestimations in the bounds of the scale of new physics $\Lambda$.


\begin{figure}[p]
    \centering
    \includegraphics[width=\linewidth]{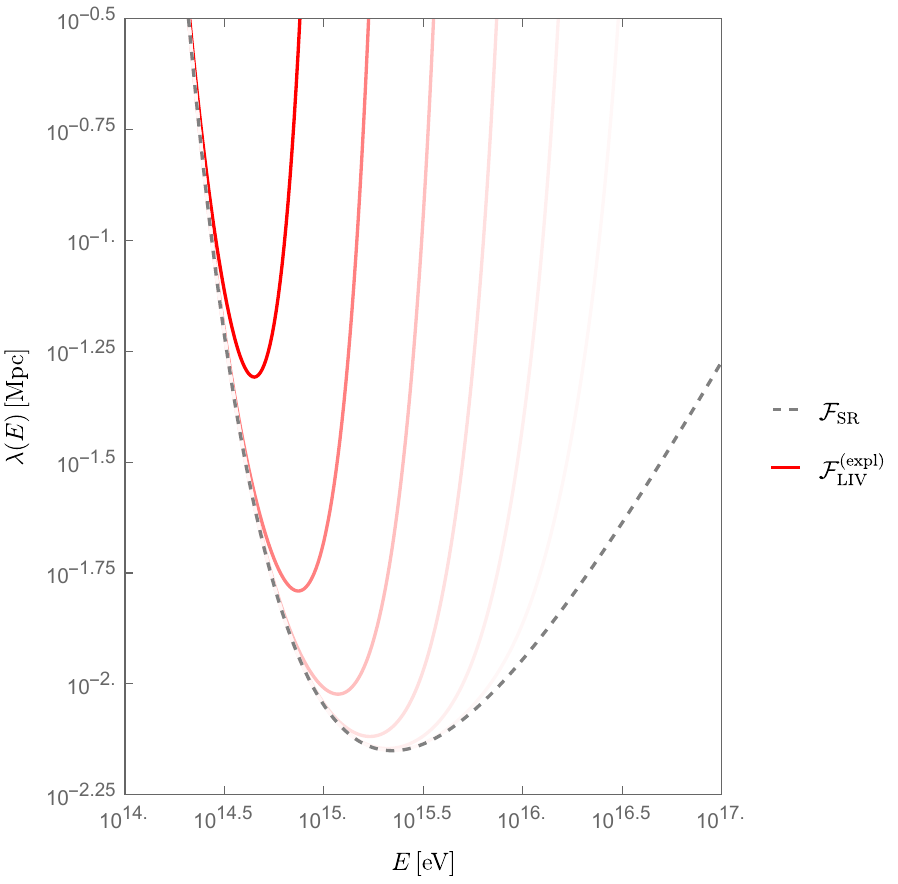}
    \caption{Mean free path for the explicit computation considering interactions with the CMB, and values of the scale of new physics $\Lambda$ (from left to right) such that $\log_{10}(\Lambda/E_\text{Pl})=-4.5$ to $-2.0$ in steps of $0.5$. The grey dashed line represents the SR (Breit-Wheeler) case.}
    \label{fig:complete_path}
\end{figure}
\begin{figure}[p]
    \centering
    \includegraphics[width=\linewidth]{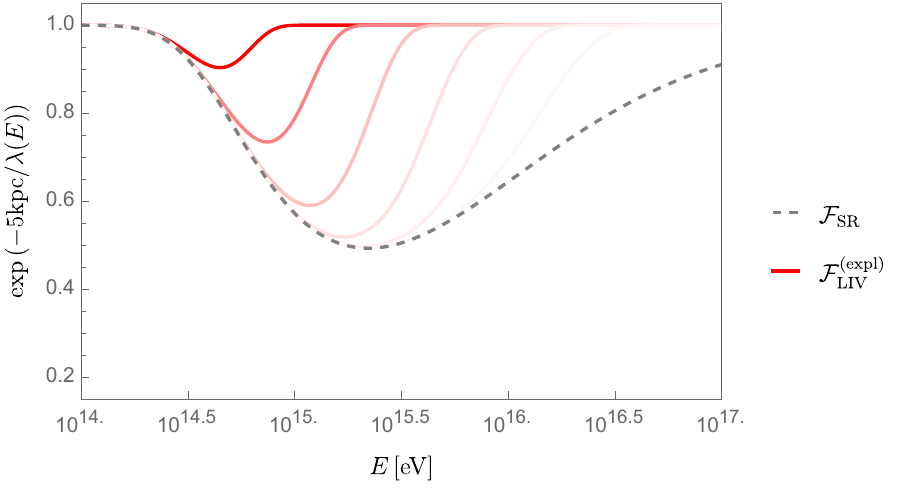}
    \includegraphics[width=\linewidth]{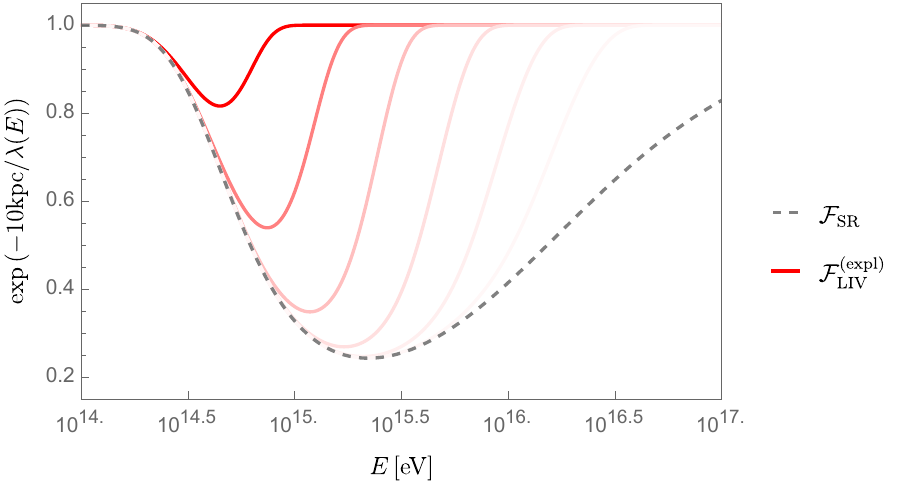}
    \caption{Survival probability for the explicit computation considering interactions with the CMB, values of the scale of new physics $\Lambda$ (from left to right) such that $\log_{10}(\Lambda/E_\text{Pl})=-4.5$ to $-2.0$ in steps of $0.5$, and two source distances, $d_s=5\mathrm{kpc}$ (upper plot) and $10\mathrm{kpc}$ (bottom plot). The grey dashed line represents the SR (Breit-Wheeler) case.}
    \label{fig:complete_prob}
\end{figure}

In order to check the previous statement, we used the following simple method. We start from hypothetical observations up to an energy of 1 PeV from a certain source, and three cases of possible experimental uncertainties $\varepsilon$ at this energy, namely 30\%, 40\% and 50\% (this includes possible statistical uncertainties in the measured flux as well as systematic uncertainties\footnote{Here we also neglect the galactic infrared background~\cite{Vernetto:2016alq}, which could affect the total absorption of photons within the Milky Way, again increasing the overall uncertainty.}). Assuming that no discrepancies with respect to SR have been found up to 1 PeV, we search for the scale $\Lambda$ at which LIV effects deviate from the SR predictions by $1\,\sigma$. 
With this procedure, we obtain lower bounds for the scale $\Lambda$ for the approximation in Eq.~\eqref{approx2} and the explicit approach in Eq.~\eqref{eq:F_complete}, which we call $\Lambda^{(2)}$ and $\Lambda^{(\text{expl.})}$ respectively.  We have repeated the analysis for two choices of distances, $d_s=5$ kpc (Table~\ref{tab:comparison_5kpc}) and $d_s=10$ kpc (Table~\ref{tab:comparison_10kpc}). We find that the use of the standard approach Eq.~\eqref{approx2} produces a 20\%-30\% underestimate in the bounds on the LIV scale with respect to the explicit calculation of Eq.~\eqref{eq:F_complete}.

\begin{table}[h]
    \setlength{\tabcolsep}{6pt}
    \centering
    \caption{Lower bounds for $\Lambda$ from the approximation \eqref{approx2} and the explicit approach \eqref{eq:F_complete}, assuming hypothetical observations at 1\,PeV with uncertainties $\varepsilon$, from a source at $d_s=5$ kpc, see text.}
    \vspace{0.5em}
    \begin{tabular}{cccc}
        \hline
        $\varepsilon$ & 30\% & 40\% & 50\% \\ \hline 
        $\Lambda^{(2)}/E_{\text{Pl}}$& $8.7\E{-5}$ & $7.5\E{-5}$ & $6.3\E{-5}$ \\ 
        $\Lambda^{(\text{expl.})}/E_{\text{Pl}}$& $1.2\E{-4}$ & $9.8\E{-5}$ & $8.1\E{-5}$ \\ 
        \hline
    \end{tabular}
    \label{tab:comparison_5kpc}
\end{table}

\begin{table}[h]
    \setlength{\tabcolsep}{6pt}
    \centering
    \caption{The same as in Table~\ref{tab:comparison_5kpc}, for a source at a distance $d_s=10$ kpc.}
    \vspace{0.5em}
    \begin{tabular}{cccc}
        \hline
        $\varepsilon$ & 30\% & 40\% & 50\% \\ \hline 
        $\Lambda^{(2)}/E_{\text{Pl}}$& $1.3\E{-4}$ & $1.1\E{-4}$ & $1.0\E{-4}$ \\ 
        $\Lambda^{(\text{expl.})}/E_{\text{Pl}}$& $1.8\E{-4}$ & $1.6\E{-4}$ & $1.4\E{-4}$ \\ 
        \hline
    \end{tabular}
    \label{tab:comparison_10kpc}
\end{table}

Using the explicit result, we can now study the effects of the subluminal quadratic LIV scenario in the mean free path and survival probability of the gamma rays, for different values of the scale of new physics. We show the results in Figs.~\ref{fig:complete_path} and \ref{fig:complete_prob}, respectively.

\subsection{Extragalactic sources}

An analogous analysis has been performed for sources outside of the Milky Way and for gamma-ray observations at 10 TeV. In this scenario one cannot disregard the redshift dependencies in Eq.~\eqref{eq:opa}. The first integral of Eq.~\eqref{eq:opa} is the distance travelled by a photon per unit of redshift
\begin{align}\label{eq:dl/dz}
    \frac{dl}{dz}\,&=\,\frac{dl}{dt}\frac{dt}{dz} \approx \frac{1}{(1+z)H(z)}\,,
\end{align}
where we have disregarded corrections $(E/\Lambda)^2$ in the velocity of the photon (as was done for the kinematical prefactor $1/\mathcal{K}$ in the cross section), and $H(z)$ is the redshift-dependent Hubble parameter. In the $\Lambda$CDM cosmological model it equals to
\begin{equation}\label{eq:lcdm}
    H(z)=H_0\sqrt{\Omega_\text{m}(1+z)^3+\Omega_\Lambda}\equiv H_0 \,h(z)\,,
\end{equation}
where we will use $\Omega_\text{m}=0.3$ and $\Omega_\text{$\Lambda$}=0.7$, as the matter and vacuum energy densities, respectively, and $H_0=70$ km s$^{-1}$ Mpc$^{-1}$ as the present value of the Hubble constant\footnote{There is a discrepancy between the value of the Hubble constant derived from direct measurements using Cepheid variables, leading to $H_0=73.04\pm1.04 \,\mathrm{km\,s^{-1}\, Mpc^{-1}}$~\cite{Riess:2021jrx}, and the value derived from the $\Lambda$CDM model and the Cosmic Microwave Background observations, $H_0=67.4\pm0.5 \,\mathrm{km\,s^{-1}\, Mpc^{-1}}$~\cite{Planck:2018vyg}. We used the fiducial value $H_0=70$ km s$^{-1}$ Mpc$^{-1}$ to avoid entering the discussion of the ``Hubble tension'', which is beyond the scope of this study. This is the same value assumed in~\cite{Saldana-Lopez:2020qzx} for the determination of the EBL model used in this work.}.

Additionally, for energies around 10 TeV, one should consider the additional contribution of the EBL to the universe transparency. Then, the total background under consideration is now
\begin{equation}
    n(\omega,z)=n_{\text{CMB}}(\omega,z)+n_{\text{EBL}}(\omega,z)\,.
\end{equation}
The redshift dependence of the CMB can be trivially accounted for by including the redshift of the temperature $T(z)=(1+z)T_0$. However, unlike $n_{\text{CMB}}$ in Eq.~\eqref{eq:ncmb}, there is no analytical formula for $n_{\text{EBL}}$ as a function of redshift. One has to use a particular model, which is in turn based on a cosmological model. Here we use the EBL model of Saldana-Lopez et al. (2021)~\cite{Saldana-Lopez:2020qzx}, in which $\Lambda$CDM cosmology is assumed, with the values for the cosmological parameters given below Eq.~\eqref{eq:lcdm}.

Introducing now the change of variables from $(\theta,\omega)$ to $(\bar\tau,\omega)$, one can rewrite the integral Eq.~\eqref{eq:opa} as
\begin{align}\label{eq:opa2}
    &\tau(E,z_s) \,=\,  \frac{m_e^2}{4E^2H_0} \int_{0}^{z_s} \frac{dz}{(1+z)^3h(z)} \notag\\ &\times \int_{1}^{\infty} d\bar\tau \,\mathcal{F}(\bar\tau,(1+z)^4\bar\mu) \int_{\omega_\text{th}(\bar\tau,z,\bar\mu)}^\infty d\omega \; \frac{n(\omega,z)}{\omega^2}\,,
\end{align}
where now
\begin{equation}
    \omega_{\text{th}}(\bar\tau,z,\bar\mu)=\frac{m_e^2}{(1+z)E}\;(\bar \tau + (1+z)^4\bar\mu)\,,\  \text{with }\ \bar\mu=\frac{E^4}{4 m_e^2 \Lambda^2}.
\end{equation} 

\begin{figure}[p]
    \centering
    \includegraphics[width=0.98\linewidth]{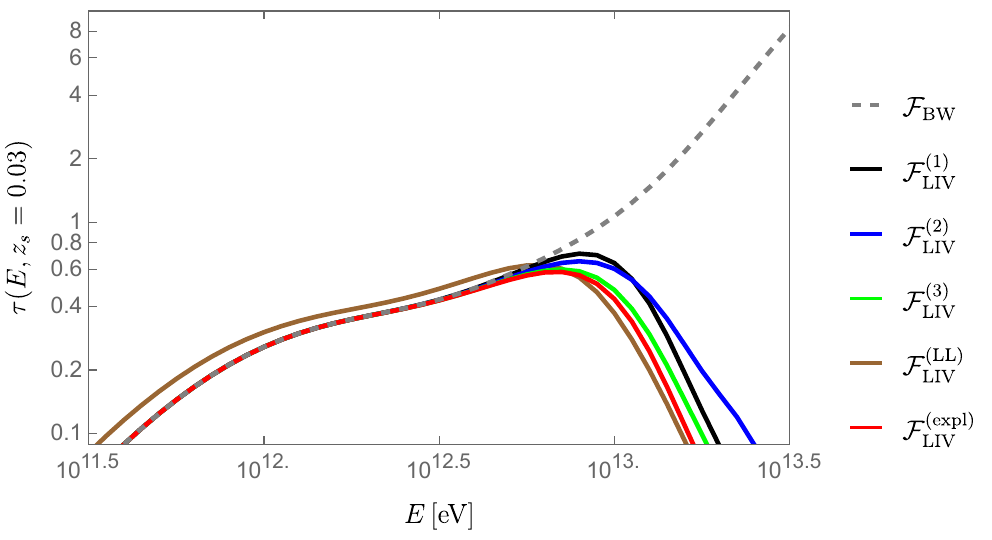}
    \includegraphics[width=0.98\linewidth]{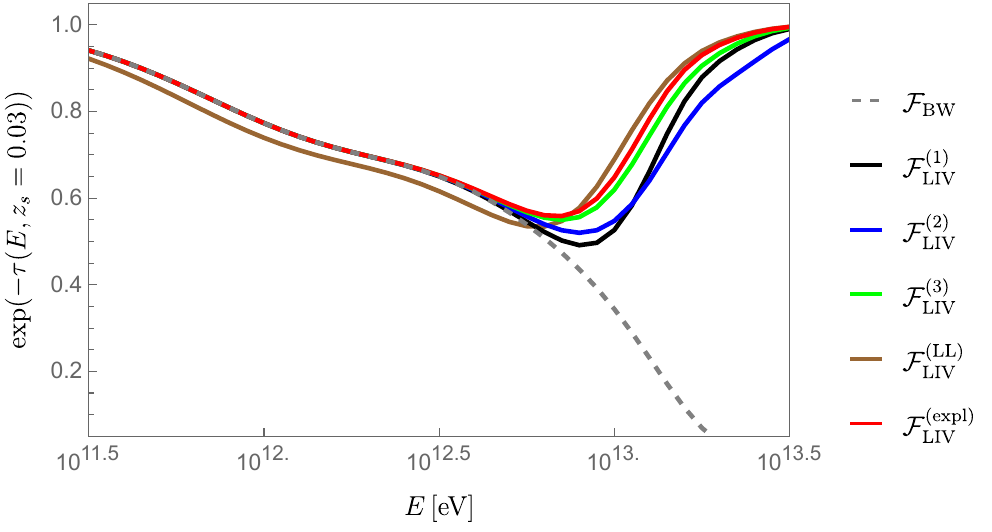}
    \caption{Comparison of the opacity (upper plot) and probability of survival (bottom plot) for the explicit (red) and leading-log computations (brown), and the three approximations (black, blue and green). We consider interactions with the CMB and EBL, $\Lambda/E_\text{Pl}=10^{-8}$, and $z_s=0.03$.}
    \label{fig:comparison_003}
\end{figure}

\begin{figure}[p]
    \centering
    \includegraphics[width=0.98\linewidth]{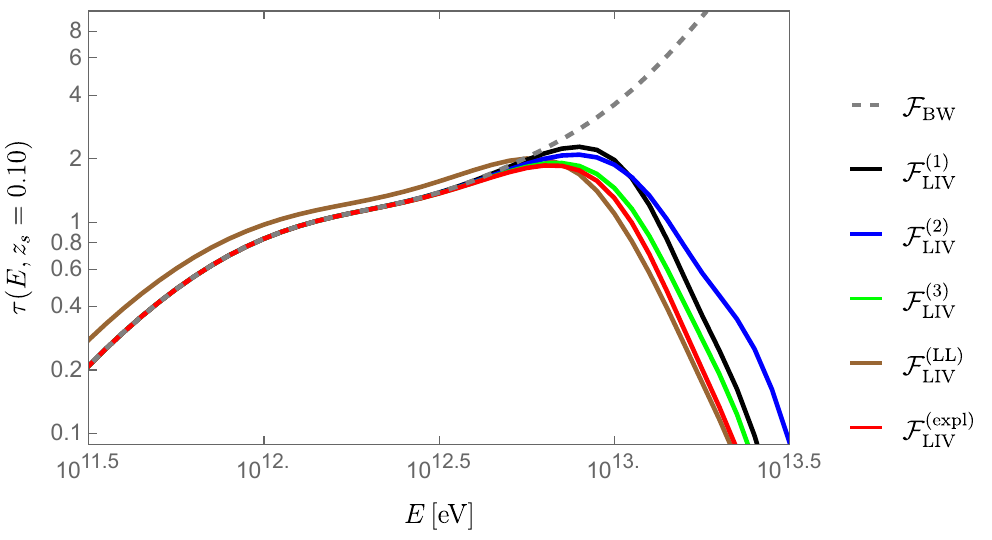}
    \includegraphics[width=0.98\linewidth]{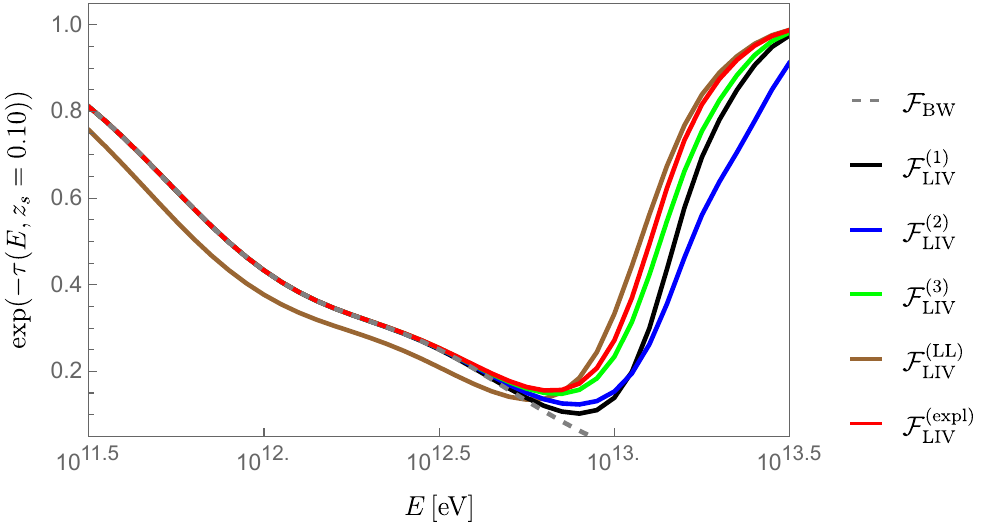}
    \caption{Comparison of the opacity (upper plot) and probability of survival (bottom plot) for the explicit (red) and leading-log computations (brown), and the three approximations (black, blue and green). We consider interactions with the CMB and EBL, $\Lambda/E_\text{Pl}=10^{-8}$, and $z_s=0.10$.}
    \label{fig:comparision_010}
\end{figure}

We show a comparison of the opacity and survival probability, for a source at redshift $z_s=0.03$ in Fig.~\ref{fig:comparison_003} and at $z_s=0.10$ in Fig.~\ref{fig:comparision_010}, using the three different approximations introduced in Section~\ref{sec:efective}, Eqs.~\eqref{approx1}, \eqref{approx2} and \eqref{approx3}, and the leading-log and explicit results of Section~\ref{sec:explicit}, Eqs.~\eqref{eq:FLL} and \eqref{eq:F_complete}.

\begin{figure}[p]
    \centering
    \includegraphics[width=0.98\linewidth]{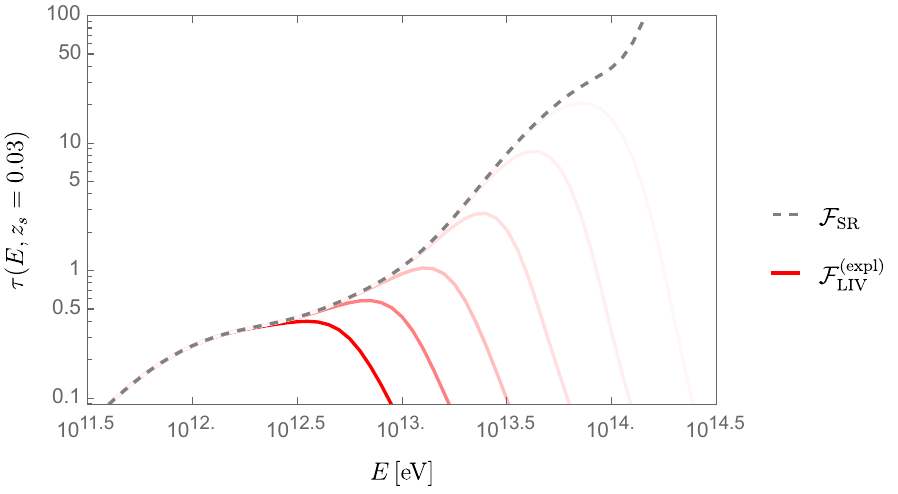}
    \includegraphics[width=0.98\linewidth]{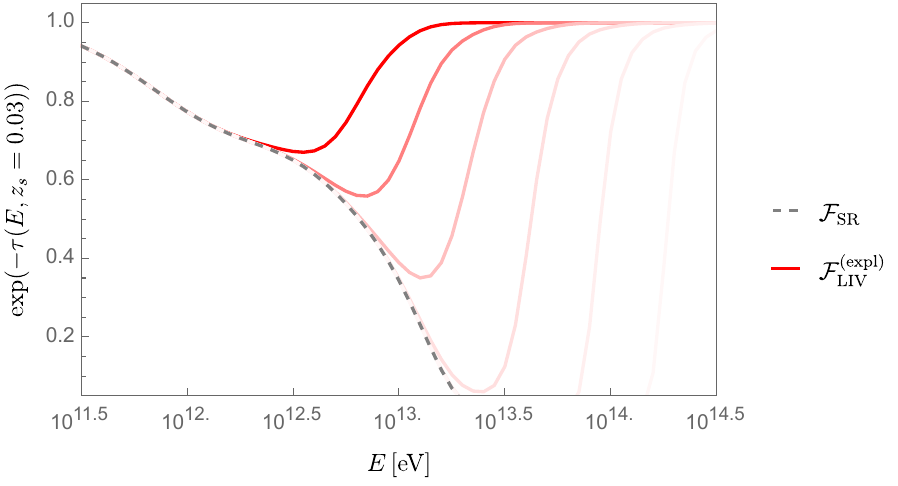}
    \caption{Opacity (upper plot) and probability of survival (bottom plot) for the explicit computation, considering interactions with the CMB and EBL, values of $\Lambda$ (from left to right) such that $\log_{10}(\Lambda/E_\text{Pl})=-8.5$ to $-6$ in steps of 0.5, and a source at $z_s=0.03$.}
    \label{fig:explicit_003}
\end{figure}
\begin{figure}[p]
    \centering
    \includegraphics[width=0.98\linewidth]{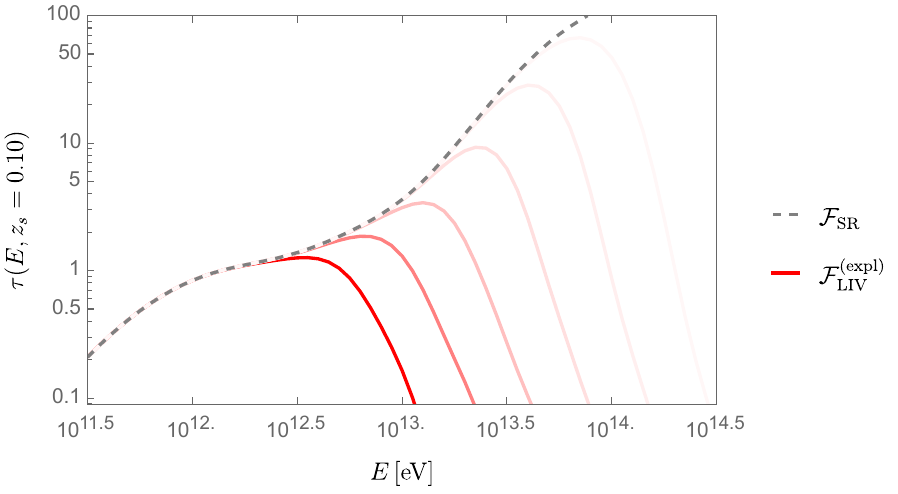}
    \includegraphics[width=0.98\linewidth]{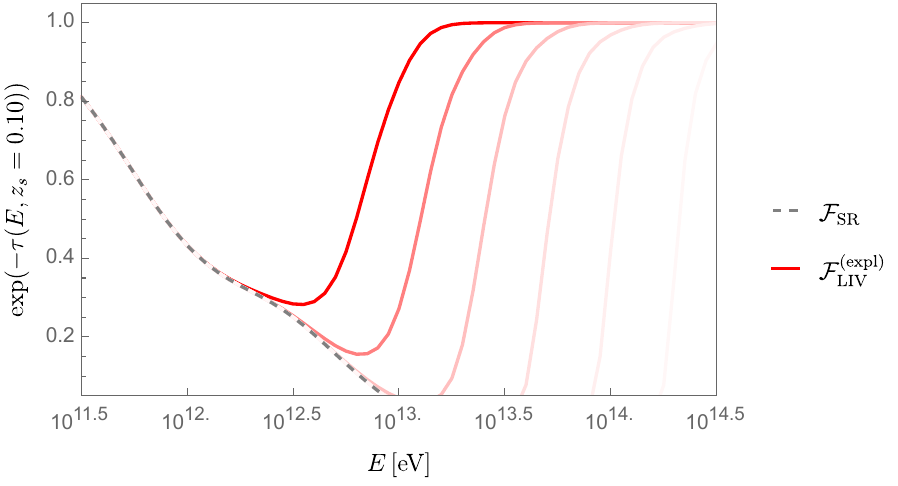}
    \caption{Opacity (upper plot) and probability of survival (bottom plot) for the explicit computation, considering interactions with the CMB and EBL, values of $\Lambda$ (from left to right) such that $\log_{10}(\Lambda/E_\text{Pl})=-8.5$ to $-6$ in steps of 0.5, and a source at $z_s=0.10$.}
    \label{fig:explicit_010}
\end{figure}

Similarly, we have done the same simple analysis presented in the previous section to obtain bounds on the scale $\Lambda$ using the approximation in Eq.~\eqref{approx2} and the explicit result of Eq.~\eqref{eq:F_complete}, but now considering no discrepancies with respect to SR up to 10 TeV. This is done for the two mentioned redshifts, $z_s=0.03$ (Table~\ref{tab:comparison_z003}) and $z_s=0.10$ (Table~\ref{tab:comparison_z01}), and assuming the same experimental uncertainties $\varepsilon= 30\%, 40\%$ and 50\%. We find that the use of the standard approach Eq.~\eqref{approx2} produces a 20\%-25\% underestimate in the bounds on the LIV scale with respect to the explicit calculation of Eq.~\eqref{eq:F_complete}.

\begin{table}[h]
    \setlength{\tabcolsep}{6pt}
    \centering
    \caption{Lower bounds for $\Lambda$ from the approximation \eqref{approx2} and the explicit approach \eqref{eq:F_complete}, coming from observations at 10\,TeV with uncertainties $\varepsilon$, from a source at a redshift $z_s=0.03$ kpc, see text.}
    \vspace{0.5em}
    \begin{tabular}{cccc}
        \hline
        $\varepsilon$ & 30\% & 40\% & 50\% \\ \hline 
        $\Lambda^{(2)}/E_{\text{Pl}}$& $1.6\E{-8}$ & $1.3\E{-8}$ & $1.1\E{-8}$ \\ 
        $\Lambda^{(\text{expl.})}/E_{\text{Pl}}$& $2.0\E{-8}$ & $1.7\E{-8}$ & $1.5\E{-8}$ \\ 
        \hline
    \end{tabular}
    \label{tab:comparison_z003}
\end{table}

\begin{table}[h]
    \setlength{\tabcolsep}{6pt}
    \centering
    \caption{The same as in Table~\ref{tab:comparison_z003}, for a source at a redshift $z_s=0.10$.}
    \vspace{0.5em}
    \begin{tabular}{cccc}
        \hline
        $\varepsilon$ & 30\% & 40\% & 50\% \\ \hline 
        $\Lambda^{(2)}/E_{\text{Pl}}$& $3.6\E{-8}$ & $3.2\E{-8}$ & $2.9\E{-8}$ \\ 
        $\Lambda^{(\text{expl.})}/E_{\text{Pl}}$& $4.5\E{-8}$ & $4.0\E{-8}$ & $3.5\E{-8}$ \\ 
        \hline
    \end{tabular}
    \label{tab:comparison_z01}
\end{table}

Using the explicit result, we can now study the effects of the subluminal quadratic LIV scenario in the mean free path and survival probability of the gamma rays, for different values of the scale of new physics. We show the results in Figs.~\ref{fig:explicit_003} and \ref{fig:explicit_010}, respectively.

\section{Conclusions}
\label{sec:conclusion}

Quantum gravity models implementing a violation of Lorentz invariance at high energies in the photon sector modify the standard expectation of the transparency of the Universe to gamma rays. In this work we focused on the particularly attractive scenario of the subluminal $n=2$ case, which is less constrained than the $n=1$ and superluminal cases, and offers a complementary view to time delay studies.

LIV introduces modifications in the physics of the interaction of gamma rays with the cosmic photon backgrounds at the level of both kinematics and dynamics. On the one hand, the apparition of an upper threshold limits the range of energies of the gamma ray that can produce electron-positron pairs in the interaction with a given background photon, leading to a kinematic suppression of pair production with respect to the SR case. On the other hand, the integration of the amplitude over the phase space produces a modification in the cross section, which, as it can be seen in Fig.~\ref{fig:comparison} for the result coming from the explicit calculation (red curves), also results in a dynamical suppression of pair production with respect to the SR case. Both types of corrections, therefore, tend to increase the transparency of the Universe to high-energy gamma rays. Even though one would naively expect corrections of order $(E/\Lambda)^2$, which would be unobservable for energies of the order of the PeV scale and values of $\Lambda$ close to the Planck scale, these corrections, coming from an effective mass for the photon $\mu^2=-E^4/\Lambda^2$ [see Eq.~\eqref{eq:mdr2}] affect the threshold of the process and can produce observable effects.

Previous studies investigating LIV modifications to the transparency of the Universe have relied on cross-section expressions derived from either speculative approximations or analytical calculations with a limited applicability range. In this work, we have provided an expression based on an explicit calculation for the $n=2$ case, Eq.~\eqref{eq:F_complete}. This expression, which has a compact and simple form, is generally valid (including situations close to the thresholds of the process) and was derived on a first-principle calculation using only the hierarchy of scales $\Lambda \gg E\gg m_e\gg \omega$. We therefore recommend researchers using this result in their phenomenological studies. In fact, we noticed that a commonly used approximation, the one indicated in Eq.~\eqref{approx2}, introduces an artificial correction in the kinematic prefactor $1/\mathcal{K}$, which produces a non-physical behaviour close to the second threshold, as it can be appreciated in Fig.~\ref{fig:comparison} (blue curves). As analyzed in Sec.~\ref{sec:transp}, the use of this approximation underestimates the effect of LIV with respect to the complete expression Eq.~\eqref{eq:F_complete}. We have evaluated this as a minor but non-negligible (around 25\%) correction. However, as more data will be available in the future, with lower uncertainties and better determination of photon fluxes at higher energies, therefore increasing the sensitivity to the high-energy scale $\Lambda$, the use of the correct expression provided in this work will become crucial to either identify or put stronger limits on LIV effects. 

Finally, we have provided in Eq.~\eqref{approx3} a new approximation to the cross section that overcomes the problems of the most common approximations used in the literature and offers a better estimate to the exact result for $n=2$. This could be particularly beneficial for investigating the $n=1$ case. The strict birefringence constraints make the $n=1$ scenario unlikely unless the description of the LIV effect extends beyond effective field theory. Given the challenges in calculating the cross-section beyond the framework of the Standard Model extension, our  approximation may offer the only viable pathway to explore such a scenario.

It should be noted that detection of VHE and UHE gamma rays relies on development of extensive air showers (EAS) in the atmosphere or water, i.e. conversion of gamma rays to electron-positron pairs in the Coulomb field of nuclei, known as the Bethe-Heitler process. A modification of the cross section by LIV can affect EAS development, thus influencing the gamma-ray detection and energy reconstruction. The modified cross section has been calculated in~\cite{Rubtsov:2012kb}, under the same assumptions as for the LIV modified Breit-Wheeler formula. An exercise analogue to the one we presented in the manuscript at hand could be repeated for the case of Bethe-Heitler process; however, that is beyond the scope of the present work and is left for a future analysis.

\FloatBarrier

\section*{Acknowledgments}
This work is supported by the Spanish grants PGC2022-126078NB-C21, funded by MCIN/AEI/ 10.13039/501100011033 and `ERDF A way of making Europe’, grant E21\_23R funded by the Aragon Government and the European Union, and the NextGenerationEU Recovery and Resilience Program on `Astrofísica y Física de Altas Energías’ CEFCA-CAPA-ITAINNOVA, and by the Croatian Science Foundation (HrZZ) Project IP-2022-10-4595, and by the University of Rijeka Project uniri-iskusni-prirod-23-24. The work of M.A.R. is supported by the FPI grant PRE2019-089024, funded by MICIU/AEI/FSE. F.R. gratefully acknowledges the Erasmus+ Mobility program of the University of Rijeka. The authors would like to acknowledge the contribution of the COST Action CA18108 ``Quantum gravity phenomenology in the multi-messenger approach''.

\bibliography{bib/Others,bib/QuGraPheno}

\end{document}